\documentclass[11pt]{article}
\usepackage{graphicx}
\usepackage{cs12,html,hyperref}

\begin{document}
 
\title{Infrared Classification of L and T Dwarfs}
 
\author{T. R. Geballe\altaffilmark{1}, G. R. Knapp\altaffilmark{2}, S. K.
Leggett\altaffilmark{3}, X. Fan\altaffilmark{4}, \\ and D. A.
Golimowski\altaffilmark{5}}
 
\altaffiltext{1}{Gemini Observatory, Hawaii}
\altaffiltext{2}{Princeton University Observatory}
\altaffiltext{3}{Joint Astronomy Centre, Hawaii}
\altaffiltext{4}{Institute for Advanced Study, Princeton}
\altaffiltext{3}{The Johns Hopkins University, Baltimore}                  

\index{classification}
\index{L dwarfs}
\index{T dwarfs}
\index{brown dwarfs}

\begin{abstract} 

We use new and published medium resolution 0.6--2.5~$\mu$m spectra of L
and T dwarfs to develop a unified classification system for both of these
new spectral classes. Two indices of the system at 1.2~$\mu$m and
1.5~$\mu$m are based on nearby absorption bands of water vapor and two are
associated with methane bands near 1.6~$\mu$m and 2.2~$\mu$m.  The
1.5~$\mu$m index is monotonic through the L and T sequences, and forms
the backbone of the system; the indices for the other bands provide
extensive, but only partial, coverage. We correlate the 1.5~$\mu$m index
with continuum indices shortward of 1~$\mu$m devised by others for
classifying L dwarfs, in order to obtain a tight link between optical and
infrared classifications. Our proposed system defines ten spectral
subclasses for L (L0--L9) and nine for T (T0--T8). The boundary between L
and T is defined to be the onset of absorption by methane in the H band.
Methane absorption in the K band near 2.2~$\mu$m is found to begin
approximately at L8.

\end{abstract}

\section{Introduction}

The recent discoveries of ultracool stellar dwarfs and brown dwarfs have
produced a need for an extension of the current system of stellar
classification, whose latest spectral class has been M since near the very
inception of classification. New spectral classes designated L and T have
been proposed and are widely used now to describe the ultracool dwarfs.
The L class (Mart\'{\i}n et al. 1999, hereafter M99; Kirkpatrick et al.
1999, hereafter K99) is generally recognized to extend from the end of the
M dwarf sequence (T$_{eff}$~$\sim$~2200~K) roughly to the temperature
($\sim$~1400~K) at which methane absorption becomes apparent in the H and
K bands, signifying that CH$_{4}$ is beginning to replace carbon monoxide
(CO) as the dominant carbon-bearing molecule. The T class (K99) includes
dwarfs with temperatures from $\sim$1400~K down at least to the coolest
($\sim$~800~K) extra-solar objects currently measured, and probably
beyond.

M99 and K99 have developed classification systems for the L sequence,
based on spectral features and continuum slopes in the optical
(0.6--1.0$\mu$m). The M99 system defines subclasses L0--L6, the K99 system
L0--L8. Using photometry and infrared (1.0--2.5~$\mu$m) spectroscopy
Leggett et al. (2000) found the first examples of dwarfs in the transition
region between the basic L and T spectral properties. Burgasser et al.
(2000) have observed a dozen T dwarfs at 1.0-2.5~$\mu$m and have
demonstrated the potential for defining a T sequence based on infrared
indices. Neither the details of a T classification scheme nor the manner
by which the two classes might be smoothly conjoined has yet been
proposed. Keys to determining these are: (1) obtaining a larger set of
infrared spectra of L and T dwarfs; (2) observing in finer detail the
transition between the classes; and (3) developing an accurate infrared
classification scheme for L dwarfs to complement the optical schemes. The
last is essential, because late L and T dwarfs cannot be easily classified
optically due to their faintness. Several groups already have begun
exploring infrared schemes for L dwarfs (Delfosse et al. 1999; Tokunaga
\& Kobayashi 1999; Reid et al. 2001; Testi et al. 2001).

This paper briefly describes a unified classification system that we have
developed for L and T dwarfs. Details are found in Geballe et al. (2001;
hereafter G01) and Leggett et al. (2001). A classification system for T
dwarfs has also recently been proposed by Burgasser et al. (2001).

\section{Observations and Analysis: Indices for 1.0--2.5~$\mu$m} 

Our work uses new and published spectra of 25 L dwarfs and 17 T dwarfs in
the 0.6-2.5~$\mu$m region. Many of these objects have been discovered in
the Sloan Digital Sky Survey. Some are 2MASS and DENIS objects classified
previously by M99, K99, and Kirkpatrick et al. (2000). Most of the spectra
were obtained by us, using CGS4 at the United Kingdom Infrared Telescope
(UKIRT) and in a few cases with NIRSPEC at the Keck~II telescope. We also
use a few spectra from Reid et al. (2001). The final reduced
1.0-2.5~$\mu$m portions of the spectra are at resolving powers near 400.
We have examined these spectra for the most useful set of infrared indices
for classification. Successful indices should be monotonic across large
swaths of the L and T sequences, based on individual spectral features or
on continuum behavior over relatively narrow wavelength regions, and
measurable at sites with average atmospheric transparency and
intermediate-sized telescopes.

\begin{figure}
\hspace{-0.3cm}
\includegraphics[width=7.0cm]{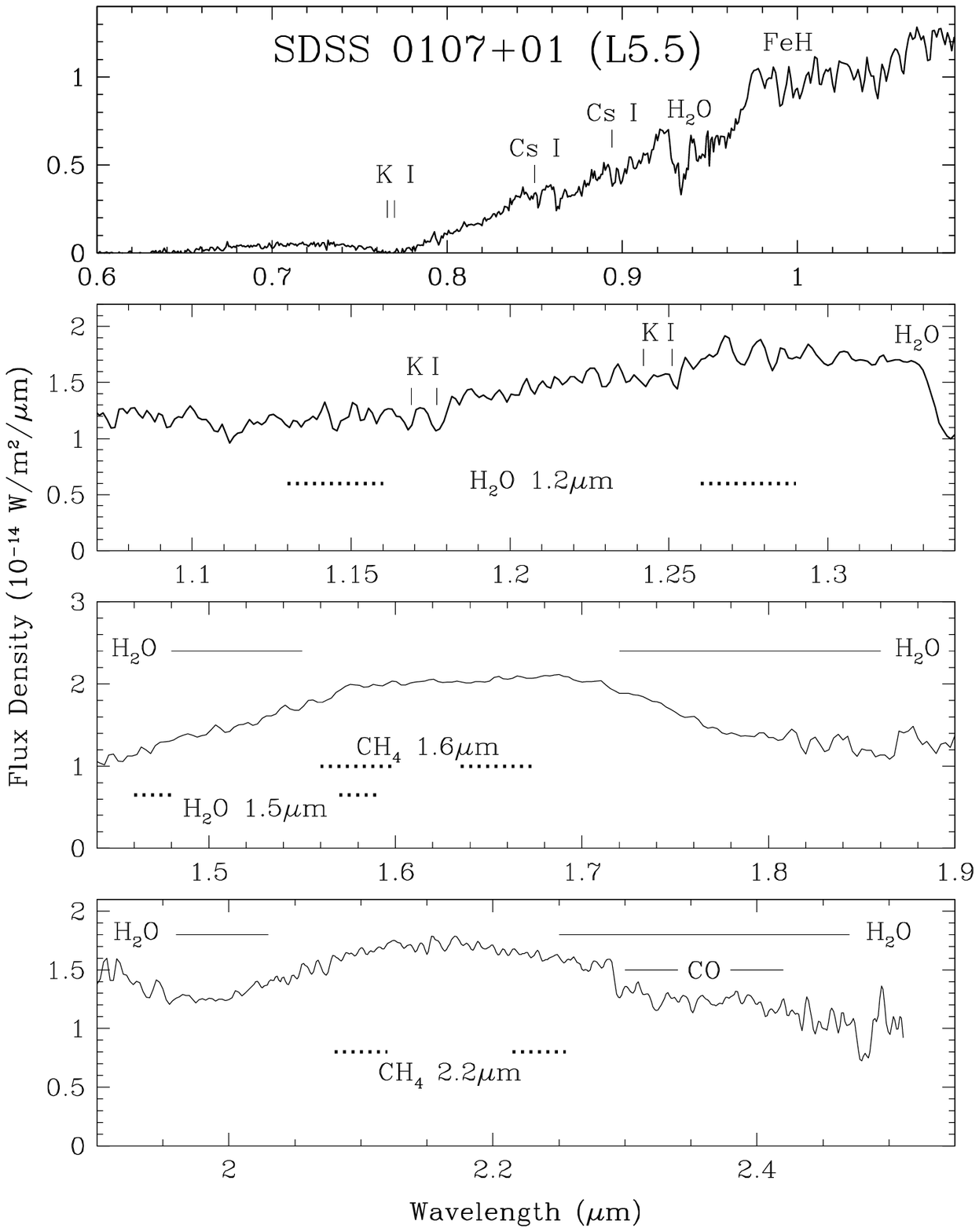}
\includegraphics[width=7.0cm]{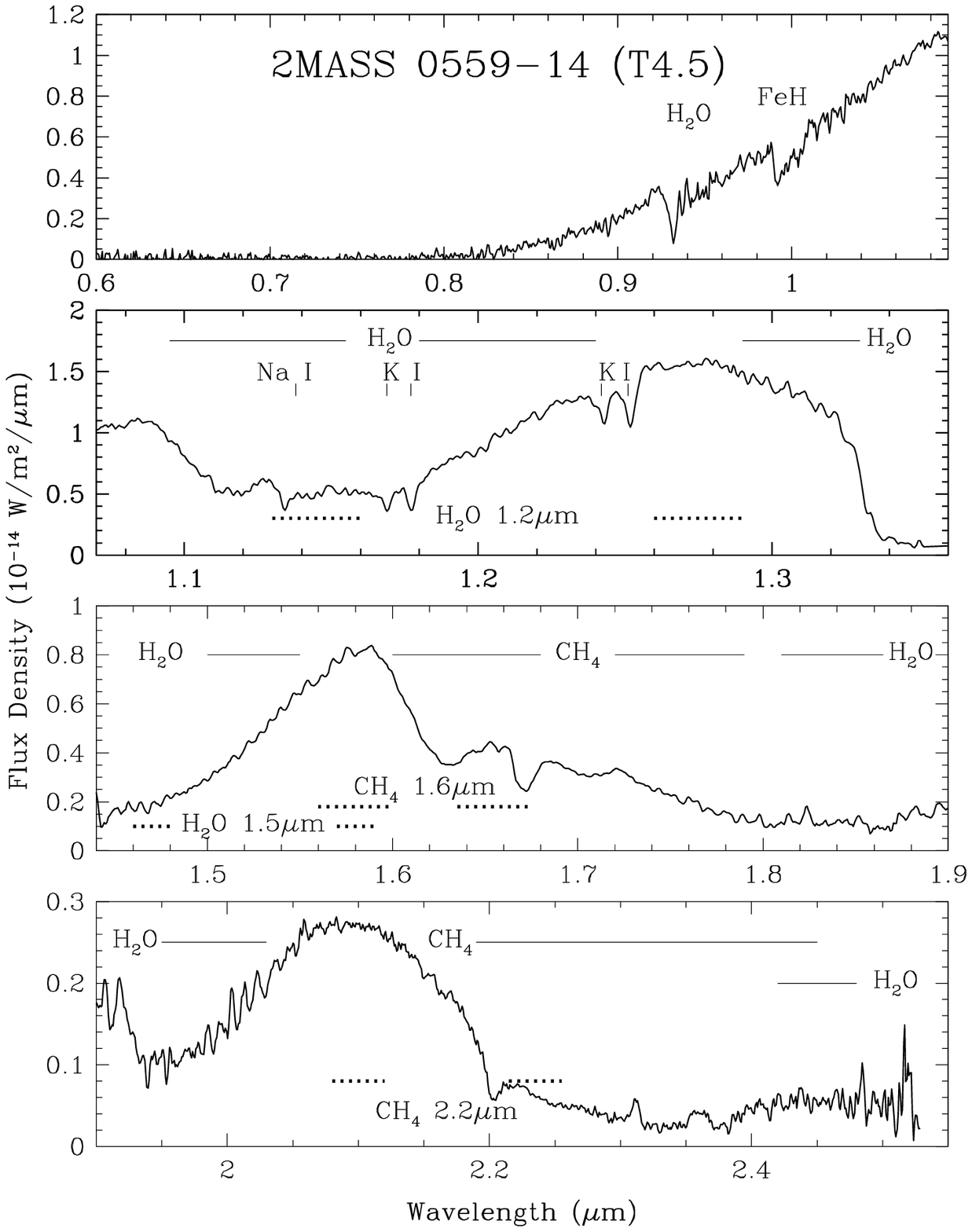}
\caption {Spectra of a mid-L dwarf and a mid-T dwarf
from 0.6 to 2.5~$\mu$m. The wavelengths of prominent spectral lines and  
bands are marked, and the wavelength ranges where the water and methane
indices are calculated are indicated by dotted lines.}
\end{figure}

We have found four indices in the 1.0--2.5~$\mu$m region that satisfy our
criteria. Two of these, near 1.15~$\mu$m and 1.5~$\mu$m, are associated
with prominent water bands and two, near 1.6~$\mu$m and 2.2~$\mu$m, with
methane bands. Their locations are shown in Fig. 1, superposed on spectra
of a mid-L and a mid-T dwarf. Each index is a ratio of fluxes in two
narrow wavelength intervals, one of which is on a portion of the molecular
absorption band and the other of which is on nearby continuum. All four
measure the increasing strengths of these bands with later spectral type.  
The 1.5~$\mu$m index is monotonic across the entire L sequence. The
2.2~$\mu$m index is useful from mid-L to the latest T dwarfs, while the
other two indices are useful only for T dwarfs. For L dwarfs we also use
two optical continuum indices ``PC3'' and ``Color-d'' from M99 and
K99, respectively (we have slightly modified the K99 index) to aid in
classification.

\begin{figure}
\hspace{-1.5cm}
\includegraphics[width=16cm]{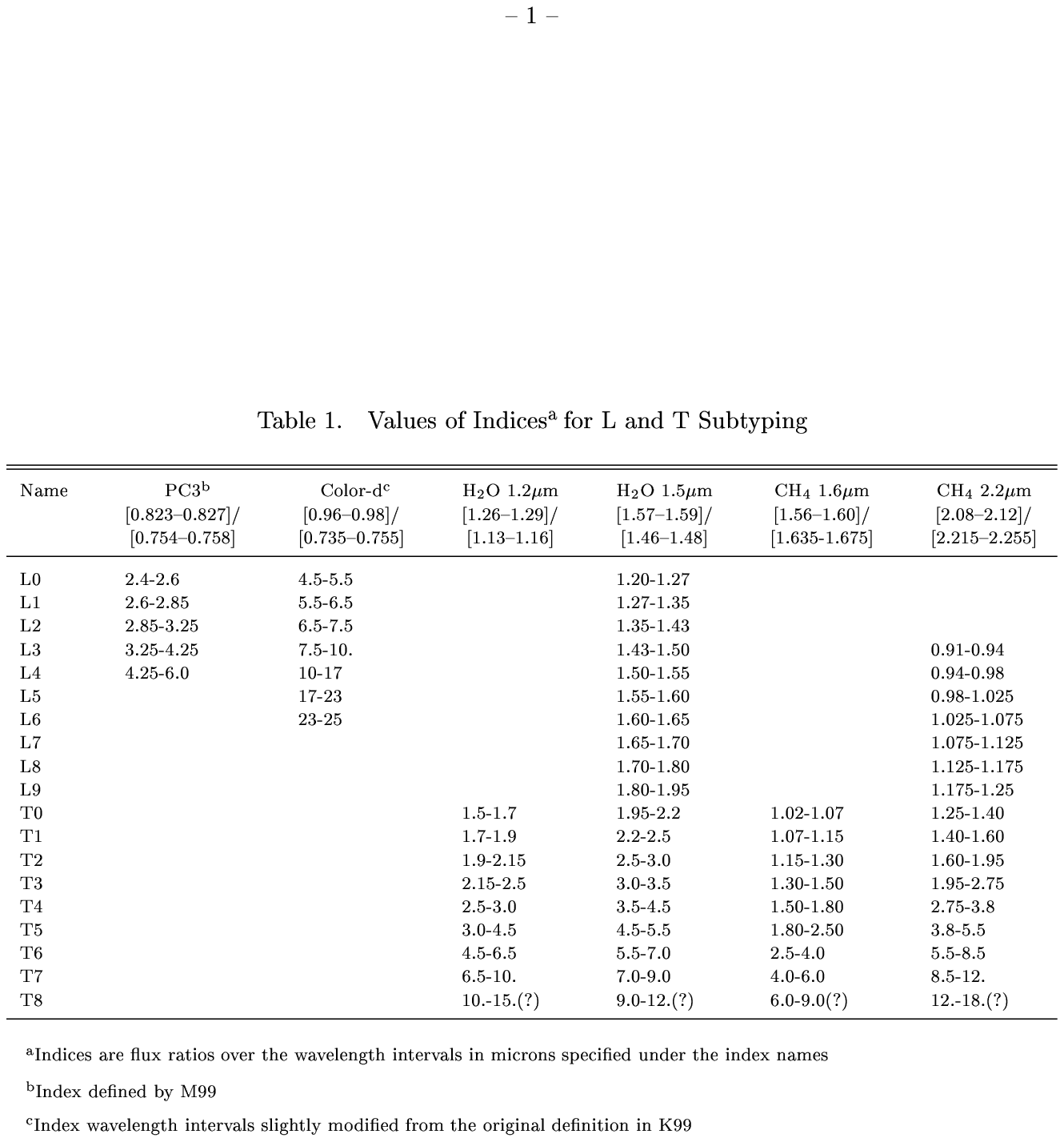}
\end{figure}

\section{The Classification System}

Table~1 lists the values of the above six indices for each
subclass. In specifying them we used the following approach.
 
\begin{itemize}

\item We defined ranges for the PC3, Color-d, and 1.5~$\mu$m indices to
make our classifications generally consistent with those of M99 and K99
for L0--L5.

\item We define a specific, and we believe sensible, phenomenological
boundary between the L and T classifications: the appearance of methane
absorption in the H band near 1.6~$\mu$m.

\item We define late L subtypes by extending the 1.5~$\mu$m and 2.2~$\mu$m
indices smoothly from L5 to this boundary, where the the 1.6~$\mu$m index 
begins to increase. This is best done by having ten L subtypes (L0--L9),
and thus our system corresponds more closely to that of K99 than M99.

\item We define T subclasses by continuing the smooth progression of the
1.5~$\mu$m and 2.2~$\mu$m indices, supplemented with 1.15~$\mu$m and
1.6~$\mu$m indices, leaving room in a system of ten subclasses for
a T9 type, yet to be discovered.

\end{itemize}

\begin{figure}
\hspace{1.0cm}
\includegraphics[width=5.45cm]{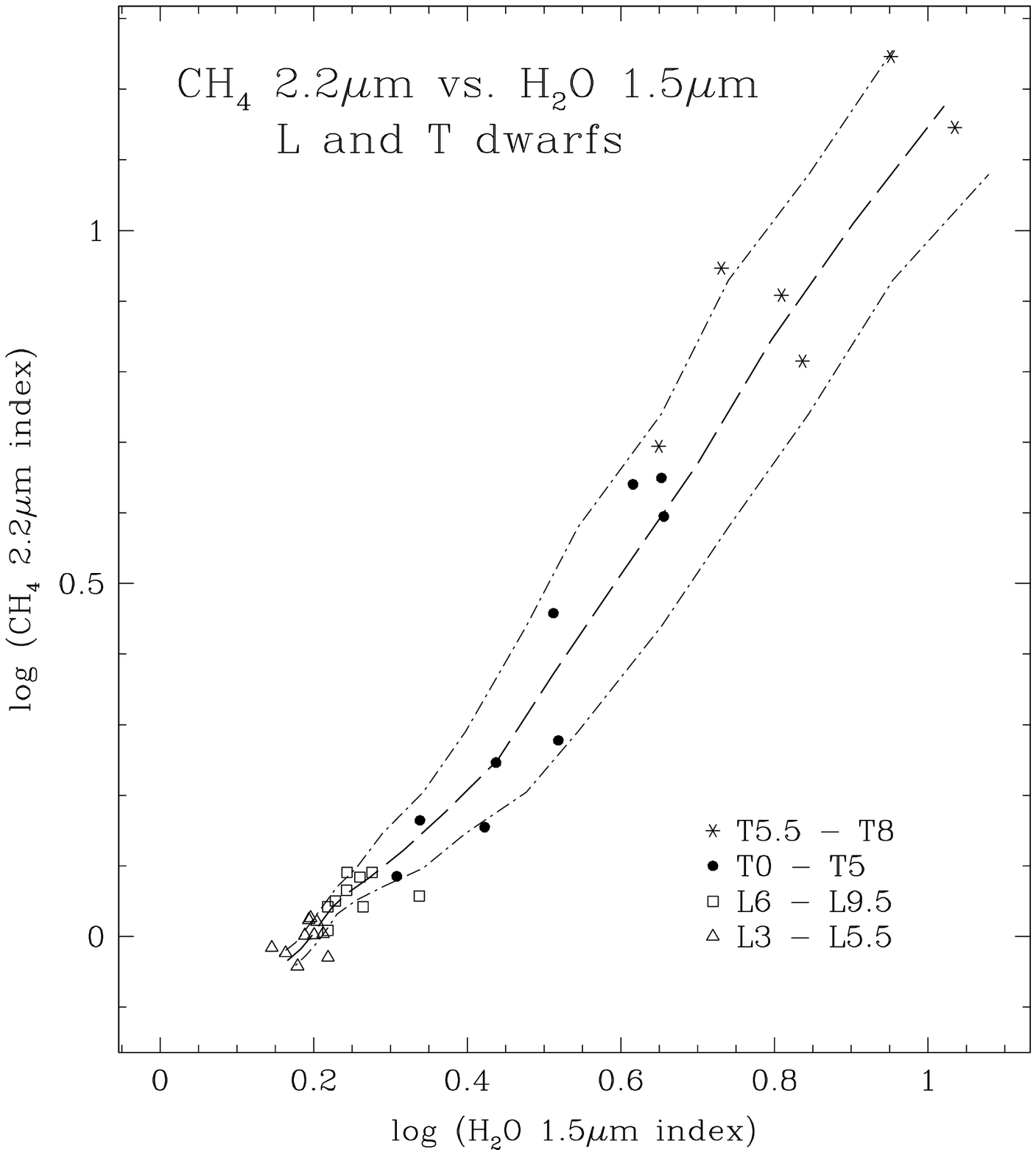}
\includegraphics[width=5.45cm]{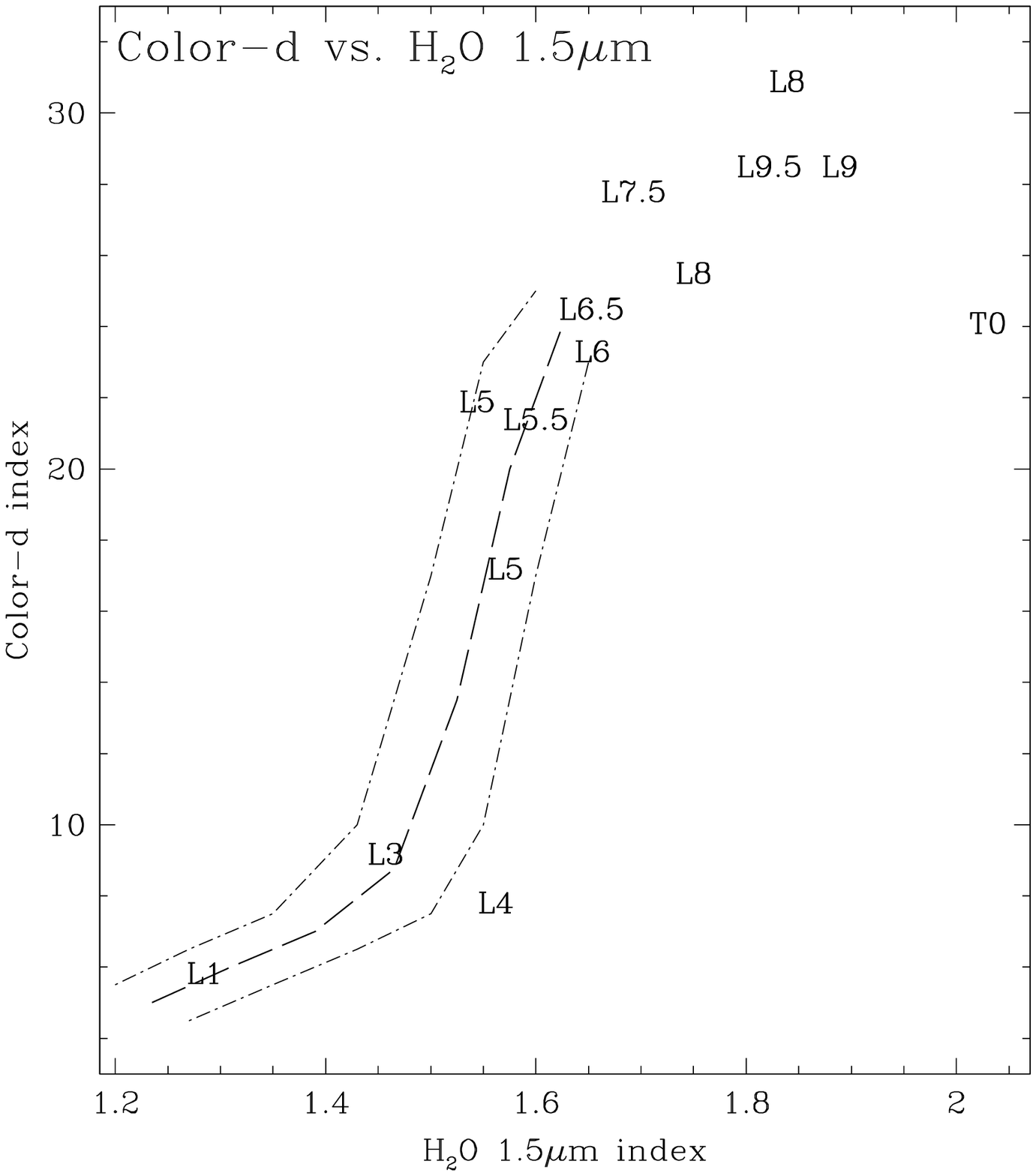}
\caption {Left-hand panel: methane 2.2~$\mu$m index plotted
against the water 1.5~$\mu$m index. The
dashed line connects the mean values for the subclasses according to the
definitions in Table~1; the dot-dashed lines differ from the mean by one
subclass. Right hand panel: modified 2MASS optical
``Color-d'' index (K99, G01) plotted against the 1.5~$\mu$m index
for L dwarfs and one T0 dwarf. The dashed line connects the
midpoints of the range of each subclass where both are defined; the
dash-dot lines deviate from the mean by one subclass. Alphanumeric symbols
are classifications of observed dwarfs.}
\end{figure}

Using this procedure we obtain self-consistent classifications for nearly
all of the L and T dwarfs in our sample. Final assignments, reported in
G01, are made by averaging the classifications from each index. Typical
uncertainties are $\pm$1 subclass for L dwarfs and $\pm$0.5 subclasses for
T dwarfs. Figure 2 illustrates the degree of self-consistency for two
pairs of indices. Although more data are required, the right hand panel
shows that the optical and infrared systems can be tightly linked in the
L0--L6 range by the Color-d and 1.5~$\mu$m indices.

\begin{figure}
\hspace{2.0cm}
\includegraphics[width=9cm]{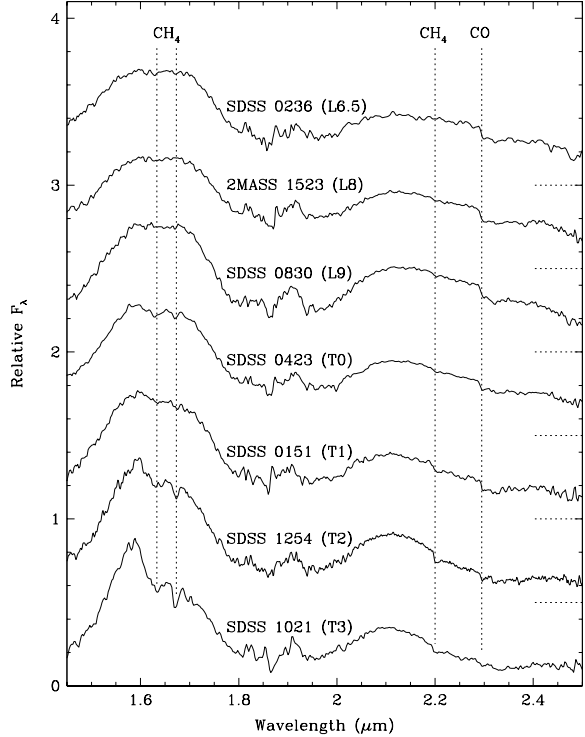}
\caption{Spectra from 1.5 to 2.5~$\mu$m of late L and early T
dwarfs, showing the onset of absorption by methane.
Horizontal dashed lines below each spectrum indicate zero flux levels.
Vertical dashed lines mark the wavelengths of spectral
features. Classifications are from G01. Spectra of
SDSS~~1021-03 and SDSS~1254-01 are from Leggett et al. (2000).}
\end{figure}

\section{The L--T boundary}

Figure 3 shows representative 1.5--2.5~$\mu$m spectra of dwarfs spanning
the L--T boundary. The 2.2~$\mu$m methane band is first evident at L8,
roughly 2 subclasses prior to the onset of H band methane absorption. The
first clear visual indication is a slight inflection at 2.20~$\mu$m due to
the sharp Q branch of the $\nu_2$+$\nu_3$ band, but the effect of the
broader absorption from the P and R branches is apparent in the index. By
T0, where the H band methane absorption first appears, the depression due
to the 2.2~$\mu$m methane band is more prominent and the CO band head at
2.29~$\mu$m has become weak. By T3 the CO band head is barely detectable
and the dominant aborber at that wavelength is the strong $\nu_3$+$\nu_4$
band, which is centered at 2.32~$\mu$m.

\section{Concluding remarks} 

In our proposed classification system the L and T sequences are seamlessly
attached via indices common to both. The system also is tied to the
optical systems of M99 and K99 via optical and infrared indices applicable
to L0--L6 dwarfs. Spectra of additional L and T dwarfs are needed to
improve the accuracy of both the subclass definitions and the
optical-infrared link. It is likely that the coolest T dwarfs currently
observed, here classified T8, are near the end of the T sequence, as all
absorption bands on which the classification is based are approaching
totality at T8 (G01).  However, once cooler dwarfs are discovered it may
be necessary to alter the definitions of the T subclasses. Finally, it is
likely that a two-dimensional classification system will be required to
incorporate the effects of surface gravity and possibly other factors such
as photospheric dust.

\acknowledgments 

UKIRT is operated by the Joint Astronomy Centre on behalf of the U.K.
Particle Physics and Astronomy Research Council. We thank the staffs of
UKIRT and the Keck Observatory for their support.  The Sloan Digital Sky
Survey (SDSS) is a joint project of the University of Chicago, Fermilab,
the Institute for Advanced Study, the Japan Participation Group, the Johns
Hopkins University, the Max Planck Institute for Astronomy (MPIA), the Max
Planck Institute for Astrophysics (MPA), New Mexico State University,
Princeton University, the United States Naval Observatory, and the
University of Washington.  Apache Point Observatory, site of the SDSS
telescopes, is operated by the Astrophysical Research Consortium (ARC).
Funding for the project has been provided by the Alfred P. Sloan
Foundation, the National Science Foundation, the U.S. Department of
Energy, the Japanese Monbukagakusho, and the Max Planck Society.  The SDSS
web site is http://www.sdss.org/

\end{document}